\documentclass[aps,pra,twocolumn,showpacs,superscriptaddress,10pt]{revtex4-1}
\usepackage[colorlinks ,linkcolor=blue,anchorcolor=blue,citecolor=blue,urlcolor=blue]{hyperref}
\usepackage{amsmath}
\usepackage{bm}
\usepackage{graphicx}
\usepackage{epstopdf}
\usepackage{tikz}
\usepackage{subfigure}
\usepackage{float} 

\usepackage{color}
\newcommand*{\blue}{\textcolor{black}}

\begin{document}

\title{Decoherence in Exchange-Coupled Quantum Spin Qubit Systems: Impact of Multiqubit Interactions and Geometric Connectivity}
\author{Quan Fu}
\affiliation{Department of Physics, City University of Hong Kong, Tat Chee Avenue, Kowloon, Hong Kong SAR, China, and City University of Hong Kong Shenzhen Research Institute, Shenzhen, Guangdong 518057, China}
\affiliation{School of Physics and Technology, Wuhan University, Wuhan
430072, China}

\author{Jiahao Wu}
\affiliation{Department of Physics, City University of Hong Kong, Tat Chee Avenue, Kowloon, Hong Kong SAR, China, and City University of Hong Kong Shenzhen Research Institute, Shenzhen, Guangdong 518057, China}

\author{Xin Wang}
\email{x.wang@cityu.edu.hk}
\affiliation{Department of Physics, City University of Hong Kong, Tat Chee Avenue, Kowloon, Hong Kong SAR, China, and City University of Hong Kong Shenzhen Research Institute, Shenzhen, Guangdong 518057, China}


\begin{abstract}
We investigate the impact of different connectivities on the decoherence time in quantum systems under quasi-static Heisenberg noise. We considered three types of elementary units, including node, stick and triangle and connect them into ring, chain, and tree configurations. We find that rings exhibit greater stability compared to chains, contrary to the expectation that higher average connectivity leads to decreased stability. Additionally, the stick configuration is more stable than the triangle configuration. We also observe similar trends in entanglement entropy and return probability, indicating their potential use in characterizing decoherence time. Our findings provide insights into the interplay between connectivity and stability in quantum systems, with implications for the design of robust quantum technologies and quantum error correction strategies.
\end{abstract}

\maketitle

\section{Introduction}

To pursue a practical quantum computer, it is necessary to successfully combine numerous qubits and ensure high levels of accuracy in single and two-qubit manipulations, as quantum error correction requires $>99.9$\% fidelity in gate operations.  Among different qubit platforms, while ultracold atoms and Si spin qubits are most promising ones for achieving scalability, operations on these systems suffer from lower fidelity rates as compared to ion traps and superconducting qubits \cite{Ladd2010,ZhangJ2017,Bloch2008,Schneider2012,Levine2019,Monroe2013,Keith2022}. While previous experimental studies have demonstrated high fidelity in operations on single \cite{YangCH2019,Gilbert2020,ChanKokWai2021} or two-qubit systems \cite{Veldhorst2014,HuangW2019,Zajac218,WatsonTF2018,XueX2019,AdamRM2022,Noiri2022,XueXiao2022}, achieving the same level of fidelity in multiqubit systems remains challenging as the scale of the system increases. Improving scalability is therefore a key problem in the study of the physical realization of quantum information processing, relevant to various platforms including dopants in semiconductors, gate-defined quantum dots, photons and atoms in cavities, Rydberg atom arrays, superconducting quantum circuits, and trapped atomic ions \cite{Britton2012,JustinG2016,Altman2021}. 

To improve the fidelity of multiqubit systems, it is important to investigate the factors that may impact their decoherence, and consequently, the fidelity of applied quantum controls as the system size increases \cite{MatthewReagor2018,fang2023,Kim2022}. For the execution of quantum algorithms, qubits must be interconnected, allowing for operations to be performed on the connected qubits. However, this connectivity often leads to decoherence, as observed in systems such as ultracold atom setups and IBM's superconducting qubits \cite{Bloch2008,ZhangJ2017}. In a quantum device, two aspects on how the qubits are connected worth consideration: the number of links between the pairs of qubits, as well as how the qubits are geometrically connected.
Intuitively, if there are less connections between qubits, the device would have longer coherence time because the qubits are less affected by their neighbors. 
Contrary to the common perception that increased connectivity invariably leads to faster decoherence, an important previous study \cite{Robert} has demonstrated that this is not always the case. The research highlighted that the geometry of the device, such as ring and chain configurations, plays a significant role in determining decoherence behavior. It was found that greater average connectivity does not necessarily result in a more rapid decay of coherence. This revelation opens up a new area of inquiry regarding the impact of geometric connectivity on decoherence times.
Inspired by this work, our research aims to provide a more comprehensive understanding of how geometric connectivity influences decoherence in multiqubit systems.

In this paper, we investigate the impact of different geometric connectivity patterns using various elementary units on the decoherence time under quasi-static Heisenberg noise. We begin by introducing a range of elementary units and then assemble them into ring, chain, and tree configurations to determine which arrangement can maintain quantum state stability for a longer duration as the system size increases. At the first glance, one may have the perception that more links between qubits leads to less stability, provided that the system size is fixed. Surprisingly, we find that the geometric connectivity plays a key role, such that the ring configuration is generally more stable than the chain, even if with a greater number of links. Furthermore, we analyze the entanglement entropy and observe similar trends to the return probability in both chain and tree configurations. These similarities suggest that both entanglement entropy and return probability can be effectively utilized to characterize the decoherence time.

The rest of the paper is organized as follows. In Sec.~\ref{Model and method}, we introduce our model and describe the elementary units and geometric connectivity configurations, including rings, chains, and trees. We also outline the method used to extract the decoherence time. Moving on to Sec.~\ref{sec:results}, we present the results of our study, comparing the decoherence time of elementary units (``node'', ``stick'', and ``triangle'') in ring and chain configurations (Sec.~\ref{subsec:result1}). Additionally, we analyze the entanglement entropy and its relationship to the return probability for ring, chain, and tree configurations (Sec.~\ref{subsec:result2}). Finally, in the concluding section, Sec.~\ref{conclusion}, we summarize our findings and their implications, highlighting the importance of geometric connectivity in understanding and mitigating decoherence effects in quantum systems.
\section{Model and method}
\label{Model and method}


In this study, we investigate the impact of geometric connectivity on decoherence time in a system with Heisenberg interaction. We consider different connectivity patterns, including ring, chain, and tree configurations constructed from multiple elementary units.

We start with a system with Heisenberg interaction on connected qubits, described by the Hamiltonian:
\begin{equation}
	H = \sum_{ij} J_{ij}(t) \boldsymbol{\sigma}_{i}\cdot\boldsymbol{\sigma}_{j}, 
	\label{H}
\end{equation}
where $ \boldsymbol{\sigma}_{i} $ is the vector of Pauli matrices on spin $ i $. Experimental evidence suggests that the Heisenberg interaction is the dominant source of noise between qubits, denoted by $J_{ij}(t)$, \blue{which is usually referred to as ``crosstalk'' in the literature \cite{Robert}}. In our analysis, we have chosen to disregard single-qubit interaction terms (such as Zeeman terms represented by $E_z \sigma_{iz}$)  to primarily focus on two-qubit crosstalk, \blue{because the errors incurred in two-qubit gates are much higher than those in single-qubit gates \cite{Johnstun2021}. In Appendix~\ref{app:single}, we have shown results that include a single-qubit term with $E_Z=0.1J_0$ (where $J_0$ is a parameter relevant to the strength of crosstalk to be described later), and have found that our main conclusions remain valid.}

By examining the evolution of decoherence time across different connectivity configurations, we aim to achieve a comprehensive understanding of how geometric connectivity influences decoherence in multiqubit systems.

Our research focuses on the evolution of a given initial state $ | \Psi(0) \rangle $ under the influence of such noise in Eq.~\eqref{H}. In other words, connected quantum bits experience crosstalk, which leads to decoherence of a prepared initial state.
We investigate how long a prepared initial state can be preserved under this noisy situation in different geometric configurations, which can be understood as undergoing a ``scrambling'' process where a closed many-body system loses memory of what it was.

In the presence of Heisenberg interaction, it is important to consider both static and dynamic aspects of the noise. In experimental settings, the frequency of noise fluctuations is much lower than the characteristic timescale of the experimental manipulations. Therefore, we primarily focus on the effect of quasi-static noise, which is characterized by a normal distribution. To account for this noise, we employ the Monte Carlo method to randomly select noise intensities and average their effects. Specifically, for quasi-static noise, we model the coupling strength $ J_{ij}(t) $ as a Gaussian distribution with a mean value $ J_{ij,0} $ and a deviation $ \sigma $. We use Monte Carlo method to realize the Gaussian distribution. For example, the interaction between qubits 1 and 2 is randomly sampled as $ J_{12}\sim N(J_{0}, \sigma^2)$ and the interaction between qubits 2 and 3 is randomly sampled as $ J_{23}\sim N(J_{0}, \sigma^2) $, etc. Note that although these $ J_{ij} $ are sampled randomly, during one Monte Carlo step they remain unchanged, implying a quasi-static noise. It is worth noting that as experimental operations become more accurate in the future, the influence of dynamic noise may become more significant. In this paper, our main focus is on the study of quasi-static noise and its impact. The effects of dynamic noise are discussed in Appendix~\ref{app:3}.

\begin{figure}[t]
	\centering
	\includegraphics[width=0.7\linewidth]{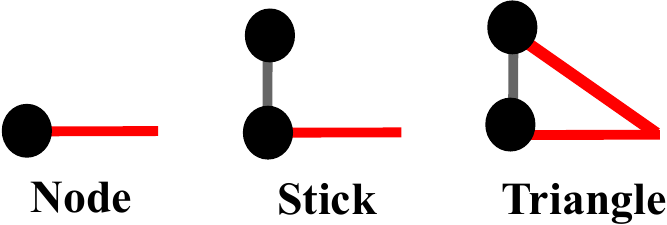}

%

	\caption{The figure illustrates the three different elementary units used in our study: ``node'', ``stick'', and ``triangle''. The vertices represent qubits, the gray lines indicate the regions within each unit where quantum gates are allowed, and the red lines represent the connections between repeated units. Our focus in this paper is to investigate the effects of different geometric connectivity configurations on the system.
	\label{fig:units}}
\end{figure}

\begin{figure}[t]
	\includegraphics[width=0.7\linewidth]{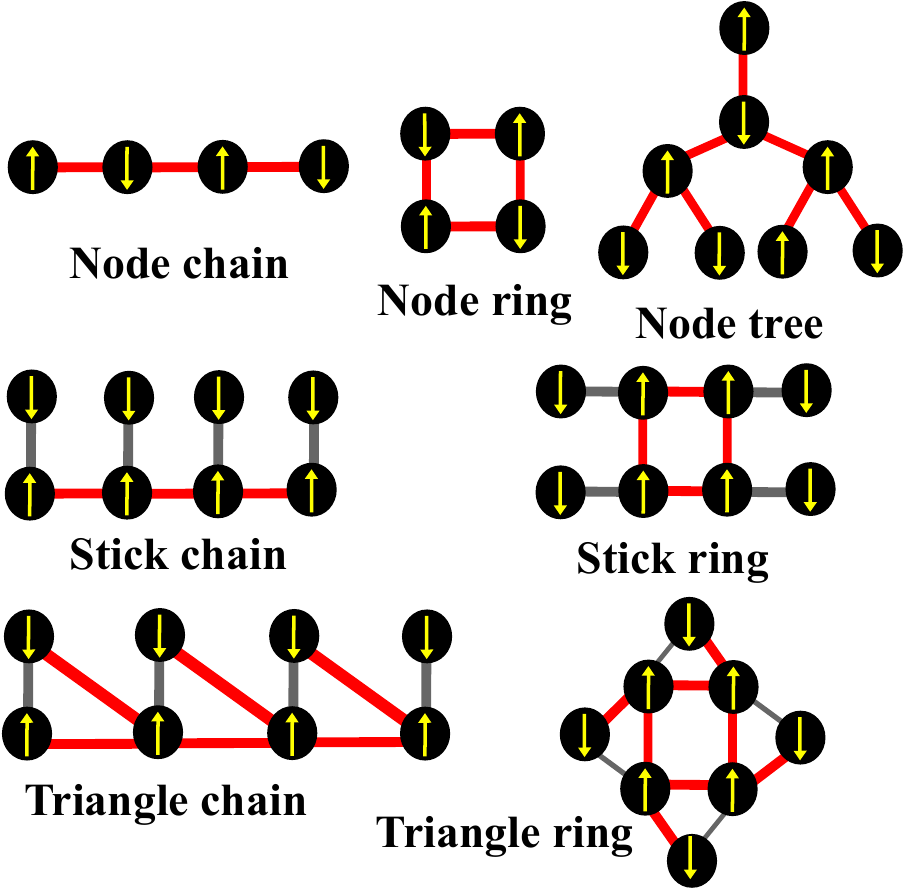}
	\caption{
		Illustration of chain, ring and tree configurations using different elementary units. The vertices represent qubits, the gray lines indicate the regions within each unit where quantum gates are allowed, and the red lines represent the connections between repeated units. The arrows inside the vertices represent the initial states that we choose. An upward arrow indicates a spin-up state at this site, while a downward arrow indicates a spin-down state. In our study, we focus on the qubits that are affected by noise-induced decoherence, which are the qubits between which quantum gates are applied. Moreover, only connected qubits can implement two-qubit gates. Therefore, we adopt a model with Heisenberg interaction between connected qubits to investigate the impact of the crosstalk noise.
		\label{fig:connect}
	}
\end{figure}

Next, we expand our model to include the study of geometric connectivity. We explore various elementary units such as the ``node'', ``stick'', and ``triangle,'' and examine their geometry in ring, chain, and tree configurations to understand how these different connectivities affect the system's behavior.
As depicted in Fig.~\ref{fig:units}, the vertices represent qubits, the gray lines represent the parts within each unit where quantum gates are allowed, and the red lines represent the connections between repeated units. These elementary units are then geometrically connected in different configurations, in the forms of a ring, chain, or tree, as shown in Fig.~\ref{fig:connect}. In this study, we investigate the decoherence time of these elementary units with $L$ ranging between 4 and 10, where $L$ represents the number of qubits. A visualization of these larger clusters are shown in Appendix~\ref{app:1}.

We extract the decoherence time by calculating the return probability
\begin{equation}
	P(t) = \left| \langle\Psi(0) | \Psi(t) \rangle \right| ^{2}, 
\end{equation}
where $ | \Psi(0) \rangle $ is the initial state and $ | \Psi(t) \rangle $ is the state at time $ t $.  We choose the initial state as $ | \Psi(0) \rangle = |\uparrow\downarrow\cdots\rangle $, shown in Fig.~\ref{fig:connect}. We note that the decoherence of a quantum system strongly depends on its initial state, as studied previously in \cite{Hung2013, Hung2014}. 
In this work, our aim is to conduct a comparative analysis of various qubit geometric configurations, which necessitates the selection of an initial state that is compatible with these different setups. The product state selected here is the most basic form that can accommodate all configurations, ensuring a fair and unbiased comparison. We have also explored the GHZ state as an alternative initial state. The findings related to this are detailed in Appendix~\ref{app:2}, and they do not alter our primary conclusion.

\begin{figure}[t]
	\includegraphics[width=0.9\linewidth]{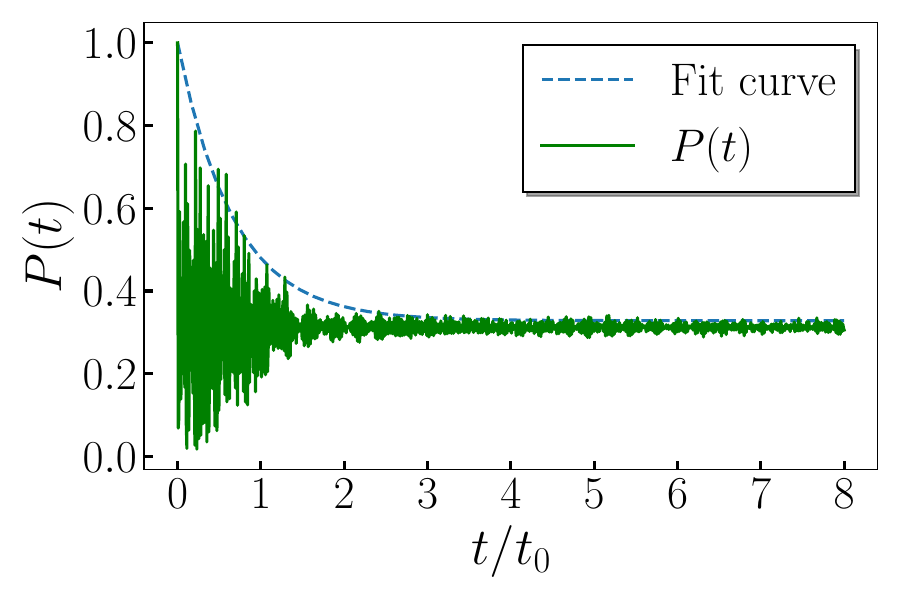}
	\caption{
		The probability $ P(t) $ as a function of time $ t $ for elementary unit ``node'', with quasi-static noise under chain configuration. The solid line represents the evolution of $P(t)$ as a function of time under $ L=4, J_0=100/t_0, \sigma = 0.4/t_0 $, $t_0$ is the time unit, while the dashed line corresponds to the fitted curve. We extract the envelope lifetime based on Eq.~\eqref{eq:evolope} to characterize the decoherence time, and we obtain $T^{*}_{2} = 0.489 t_0$. 
		\label{fig:T2}
	}
\end{figure}

The return probability $ P(t)\propto P_{A}(t)P_{\phi}(t) $ can be expressed as
\begin{equation}
	P_{A}(t) = P_{\infty} \pm (1-P_{\infty}) e^{-(t/T_{2}^{*})^{\alpha}}, 
	\label{eq:evolope}
\end{equation}
    where $P_{A}(t)$ is the amplitude term, $P_{\phi}(t)$ is the frequency term, and the decoherence time $ T^{*}_{2} $ is extracted from fitting. In the case of elementary unit ``node'' with quasi-static noise, we analyze the probability $ P(t) $ as a function of time $ t $. Fig.~\ref{fig:T2} shows the evolution of $ P(t) $ for $ L=6, J_0=100/t_0, \sigma = 0.4/t_0 $, represented by the solid line. In this work, we denote $t_0$ as our time unit, which is about 100 $ \mu $s in superconducting qubits \cite{Rigetti2012} and about 700 $ \mu $s for a typical iron trap quantum device \cite{Zhang2020}.
The dashed line corresponds to the fitted curve, while the dotted line represents the asymptotic value $ P_{\infty} $. We extract the envelope lifetime using the Eq.~\eqref{eq:evolope}, and obtain decoherence time is $ T_2^{*} = 0.489t_0 $. In the case of a transmon qubit system, $T_2^{*}$ is around 95 $ \mu $s, in consistency with experiments. 
\section{Results}
\label{sec:results}




\subsection{Decoherence time under quasi-static noise}
\label{subsec:result1}
We initially focus on the ring and chain connectivity configurations. We will discuss the tree configuration in a later stage because the system size involved is significantly larger. 
We specifically investigate the results of the ``node'' elementary unit's tree configuration for a system size of $L=8$ in Sec.~\ref{subsec:result2}. 

\begin{figure}[t]
	\includegraphics[width=0.9\linewidth]{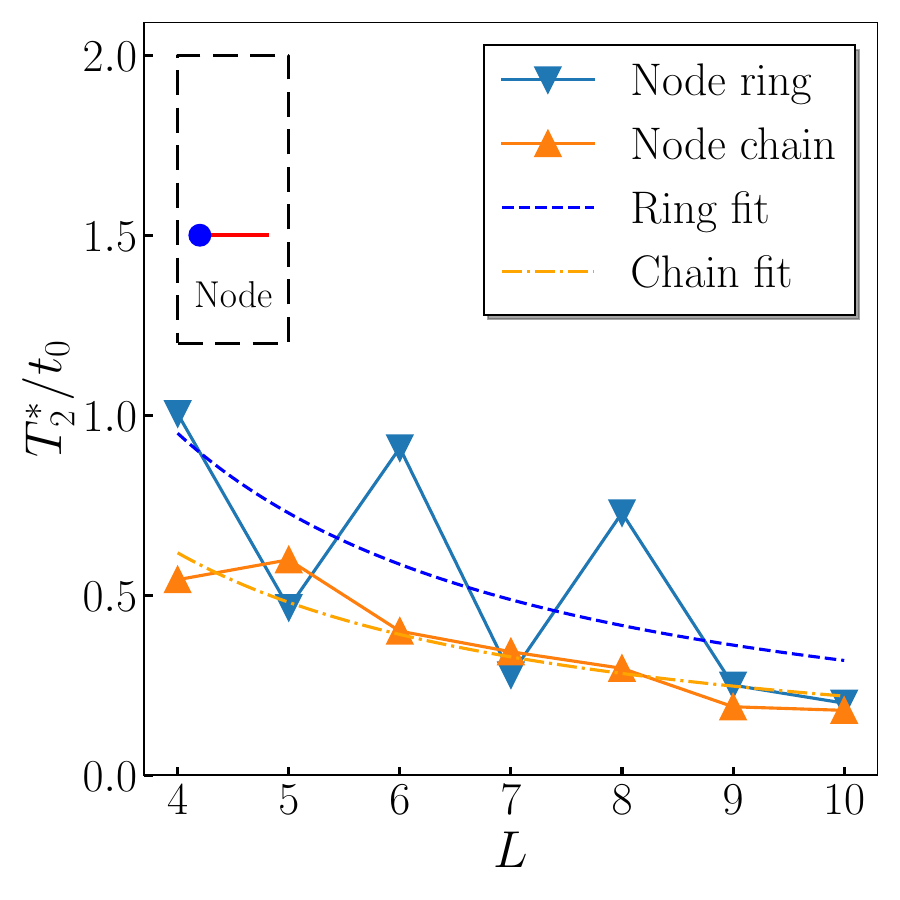}
	\caption{
		The decoherence time $ T_2^{*} $ as a function of system size $ L $ for elementary unit ``node'', with quasi-static noise. The blue inverted triangle markers correspond to the ring configuration, while the orange upright triangle markers represent the chain configuration.  \blue{The dashed line represents the fitting curve for the ring using the formula $ T_2^{*}(L)=\tau_0 L^{-\gamma} $, while the dash-dotted line represents the fitting curve for the chain.} For the ring configuration, the fitting parameters are $ (\tau_0, \gamma) $ = (4.957, 1.192), and for the chain configuration, the parameters are (2.950, 1.197). The interacting strength $ J_0=100/t_0 $ and the deviation $ \sigma=0.5/t_0 $. The decay trend is similar to Ref.~\cite{Robert}. 
		\label{fig:Ascaling}
	}
\end{figure}

Firstly, we extract the decoherence time $ T_2^{*} $ for the elementary unit ``node'', with the system size up to $ L=10 $. We set the interaction strength as $ J_0=100/t_0 $ and the deviation as $ \sigma=0.5/t_0 $. Choosing $ J_0=100/t_0 $ allows for faster decay of the system and more accurate calculation of $ T_2^{*} $, as the product of the energy unit $ J_0 $ and the decoherence time remains constant. Fig.~\ref{fig:Ascaling} shows that the decoherence time $ T_2^{*} $ exhibits similar behavior to that reported in Ref.~\cite{Robert}. We  fit these results using the power law form $ T_2^{*}(L)=\tau_0 L^{-\gamma} $, finding the corresponding fitting parameters $ (\tau_0, \gamma) $ for ring are (4.957, 1.192), and for chain are (2.950, 1.197). It can be observed that in the ring configuration, with the ``node'' as the elementary unit, the overall decoherence time decreases as the system size $L$ increases. Specifically, the decoherence time decreases from $T_2^{*}=1.004 t_0$ at $L=4$ to $T_2^{*}=0.201 t_0$ at $L=10$, resulting in a decline of approximately $80\%$. On the other hand, for the chain configuration, the decoherence time also decreases with increasing system size $L$, but at a slower rate compared to the ring configuration. Similar results have been reported in Ref.~\cite{Robert}. Specifically, for the chain, the decoherence time decreases from $T_2^{*}=0.543 t_0$ at $L=4$ to $T_2^{*}=0.297 t_0$ at $L=10$, corresponding to a decline of approximately $45\%$. Minor discrepancies between our numerical results and those of Ref.~\cite{Robert} may arise from differences in the chosen energy scale, variations in the Monte Carlo process, and numerical fluctuations when extracting $T_2^{*}$ from the envelope fitting.

In Fig.~\ref{fig:Ascaling}, we notice an oscillatory behavior in the dephasing time for the ring configuration of nodes. Specifically, there are noticeable dips for configurations with an odd number of qubits, suggesting that these configurations have a shorter lifetime compared to similar cases with an even number of qubits.
 This phenomenon can be attributed to an ``even-odd'' effect, which is linked to the number of spins in the system. When the number of spins is even, each spin can align itself in the opposite direction to its two neighbors. However, in a system with an odd number of spins, at least one spin will be unable to align oppositely to both its neighbors, resulting in a state of ``frustration.'' This frustrated state has a higher local entropy, making it less stable and leading to shorter lifetimes. This effect is specific to ring configurations with an odd number of nodes and does not occur in configurations based on sticks and triangles, where the number of qubits is always even. Similarly, chain configurations do not exhibit this frustration effect. In the analysis of GHZ states presented in Appendix~\ref{app:2}, we do not observe this even-odd effect, as expected, because GHZ states are symmetrized and thus not subject to frustration.


\begin{figure}[t]
	\includegraphics[width=0.9\linewidth]{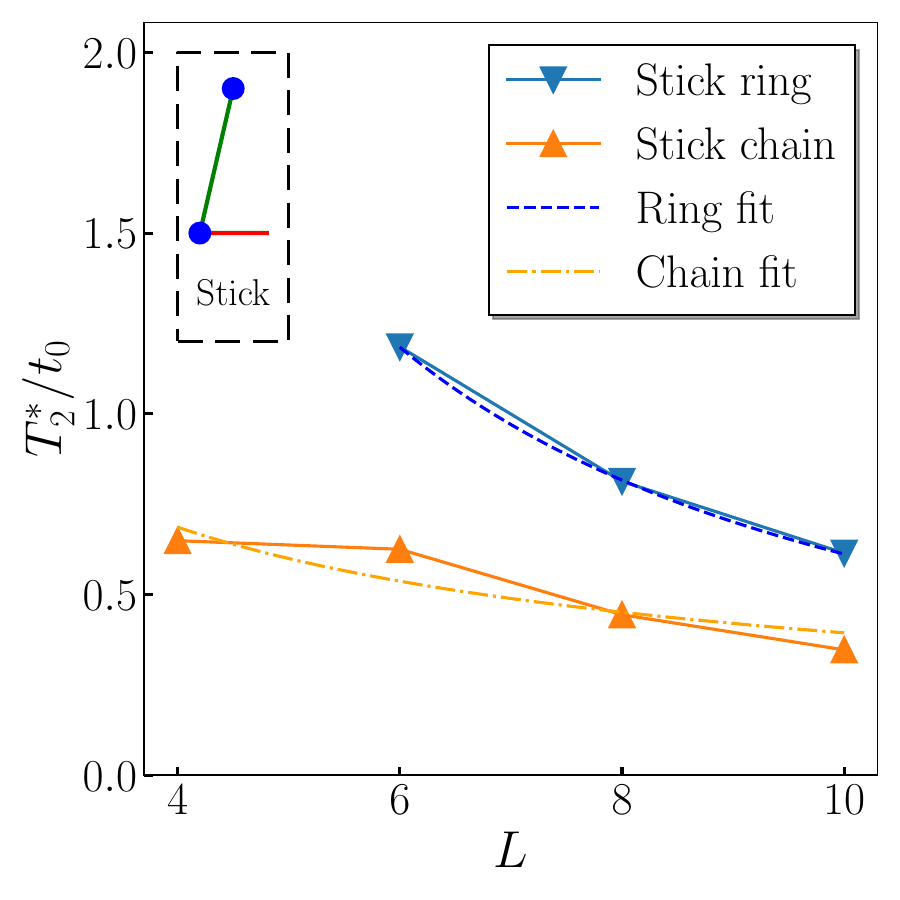}
	\caption{
		The decoherence time $ T_2^{*} $ as a function of system size $ L $ for elementary units ``stick'', with quasi-static noise. The blue inverted triangle markers correspond to the ring configuration, while the orange upright triangle markers represent the chain configuration.  \blue{The dashed line represents the fitting curve for the ring using the formula $ T_2^{*}(L)=\tau_0 L^{-\gamma} $, while the dash-dotted line represents the fitting curve for the chain.} For the ring configuration, the fitting parameters are $ (\tau_0, \gamma) $ = (12.008, 1.293), and for the chain configuration, the parameters are (1.589, 0.606). The interacting strength $ J_0=100/t_0 $ and the deviation $ \sigma=0.5/t_0 $. The decoherence time for ring configuration is longer than that of chain configuration. 
		\label{fig:Bscaling}
	}
\end{figure}

Next, we investigate the behavior of elementary units ``stick'' and ``triangle'' up to $L=10$, considering an interacting strength of $J_0=100/t_0$ and a deviation of $\sigma=0.5/t_0$. In Fig.~\ref{fig:Bscaling}, we observe that for the ring configuration with the ``stick'' as the elementary unit, the overall decoherence time decreases as the system size $L$ increases. Specifically, the decoherence time decreases by $48\%$, from $T_2^{*}=1.185 t_0$ at $L=6$ to $T_2^{*}=0.614 t_0$ at $L=10$. We note that we excluded the case of $L=4$ for the ``stick'' elementary unit, as it comprises only two elementary units, which cannot form a ring. We have fitted the results using the power law form $ T_2^{*}(L)=\tau_0 L^{-\gamma} $. The corresponding fitting parameters $ (\tau_0, \gamma) $  for the ring configuration are (12.008, 1.293), and for the chain configuration, they are (1.589, 0.606). As for the chain configuration with the ``stick'' as the elementary unit, the overall decoherence time also decreases with increasing system size $L$. Comparatively, at $L=6$, the decoherence time decreases from $T_2^{*}=0.625 t_0$ to $T_2^{*}=0.444 t_0$ at $L=10$, representing a decrease of $29\%$. In summary, the chain exhibits a slower rate of decrease in decoherence time with increasing size. However, the ring configuration consistently displays greater stability than the chain at the same system size.

\begin{figure}[t]
	\includegraphics[width=0.9\linewidth]{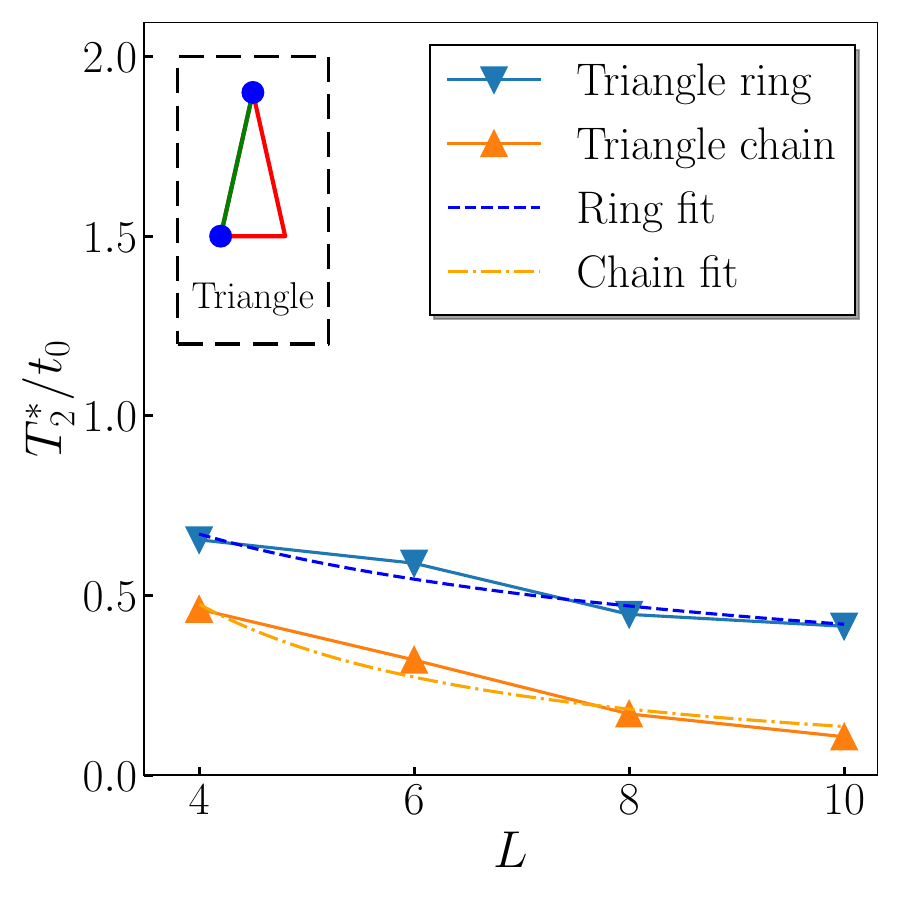}
	\caption{
		The decoherence time $ T_2^{*} $ as a function of system size $ L $ for elementary units ``triangle'', with quasi-static noise. The blue inverted triangle markers correspond to the ring configuration, while the orange upright triangle markers represent the chain \blue{The dashed line represents the fitting curve for the ring using the formula $ T_2^{*}(L)=\tau_0 L^{-\gamma} $, while the dash-dotted line represents the fitting curve for the chain.}  For the ring configuration, the fitting parameters are $ (\tau_0, \gamma) $ = (1.363, 0.511), and for the chain configuration, the parameters are (3.182, 1.371). The interacting strength $ J_0=100/t_0 $ and the deviation $ \sigma=0.5/t_0 $. The decoherence time for ring configuration is also longer than that of chain configuration. 
		\label{fig:Cscaling}
	}
\end{figure}

In Fig.~\ref{fig:Cscaling}, we observe consistent trends in the decoherence time for both ring and chain configurations. For the ring configuration with the ``triangle'' as the elementary unit, the decoherence time decreases gradually as the system size $L$ increases. Specifically, $T_2^{*}$ decreases from $0.655 t_0$ for $L=4$ to $0.414 t_0$ for $L=10$. On the other hand, in the chain configuration with the ``triangle'' as the elementary unit, the overall decoherence time also decreases with increasing system size $L$, showing a similar rate of decrease. The decoherence time decreases from $T_2^{*}=0.462 t_0$ for $L=4$ with  to $T_2^{*}=0.107 t_0$ for $L=10$. Similarly, the fitting parameters $ (\tau_0, \gamma) $ for the ring configuration are  (1.363, 0.511), , and for the chain configuration, they are (3.182, 1.371) when using the power law form $ T_2^{*}(L)=\tau_0 L^{-\gamma} $. The overall rate of decay is relatively similar between the ring and chain configurations, highlighting the distinct influence of the elementary unit ``triangle'' and ``stick'' on the decoherence dynamics.

Intuitively, one might expect that a lower average number of qubit connections would extend the decoherence time. The reasoning is straightforward: fewer qubit connections mean fewer positions for applying quantum gates, which could potentially reduce the impact of noise on the system. However, in Fig.~\ref{fig:Bscaling} and Fig.~\ref{fig:Cscaling}, we observe that the ring configuration is generally more stable than the chain configuration, despite the fact that the average number of connections between qubits in a ring consistently exceeds that in a chain. This counterintuitive relationship between the complexity of qubit interconnections and decoherence dynamics could provide valuable insights into optimizing quantum computing architectures for improved stability.
We also observe that the decoherence times for configurations formed by elementary units ``stick`` and ``triangle`` do not exhibit the oscillatory dependence seen in Fig.~\ref{fig:Ascaling}. This absence of oscillatory behavior is due to the fact that these configurations always contain an even number of qubits, eliminating the ``even-odd'' effect.

Now, we proceed to calculate the decoherence time $ T_2^{*} $ as a function of the system size $ L $ for the elementary units ``stick'' and ``triangle'' with ring and chain configurations, considering different deviation values $ \sigma $. 
\begin{figure}[t]
	\centering
	\setcounter{subfigure}{0}(a){
		\includegraphics[width=0.42\linewidth]{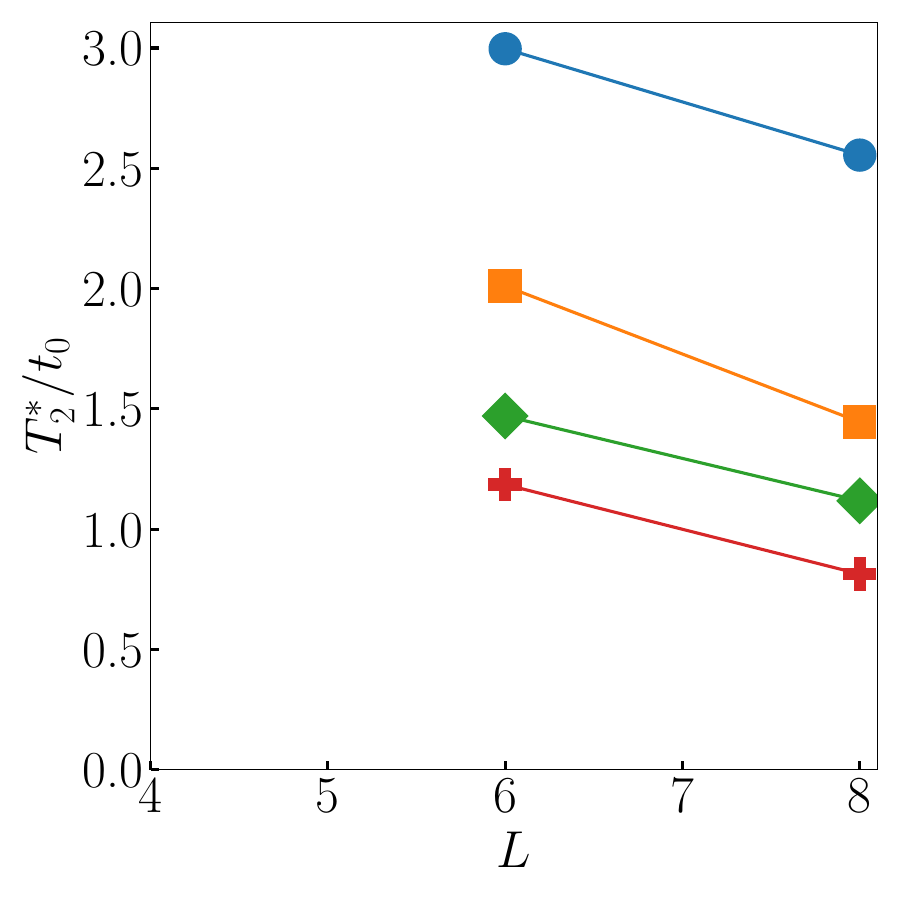}
		\label{fig:subfig_a}}
	\hfill
	\setcounter{subfigure}{1}(b){
		\includegraphics[width=0.42\linewidth]{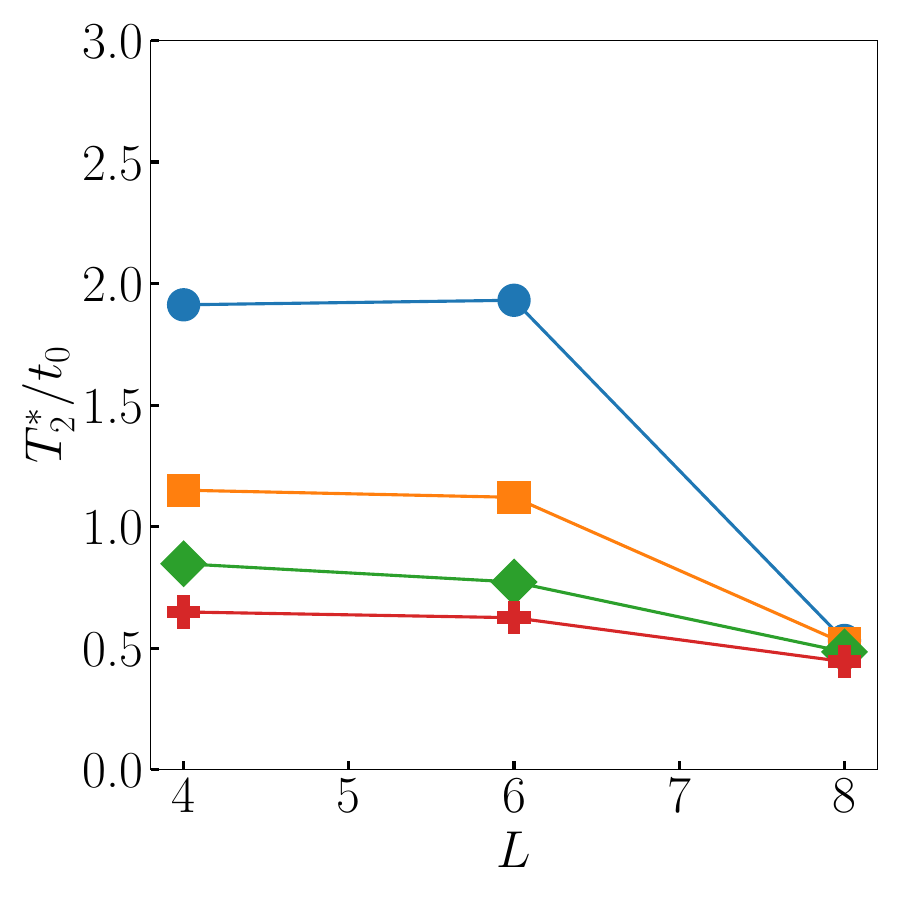}
		\label{fig:subfig_b}}
	
	\setcounter{subfigure}{2}(c){
		\includegraphics[width=0.42\linewidth]{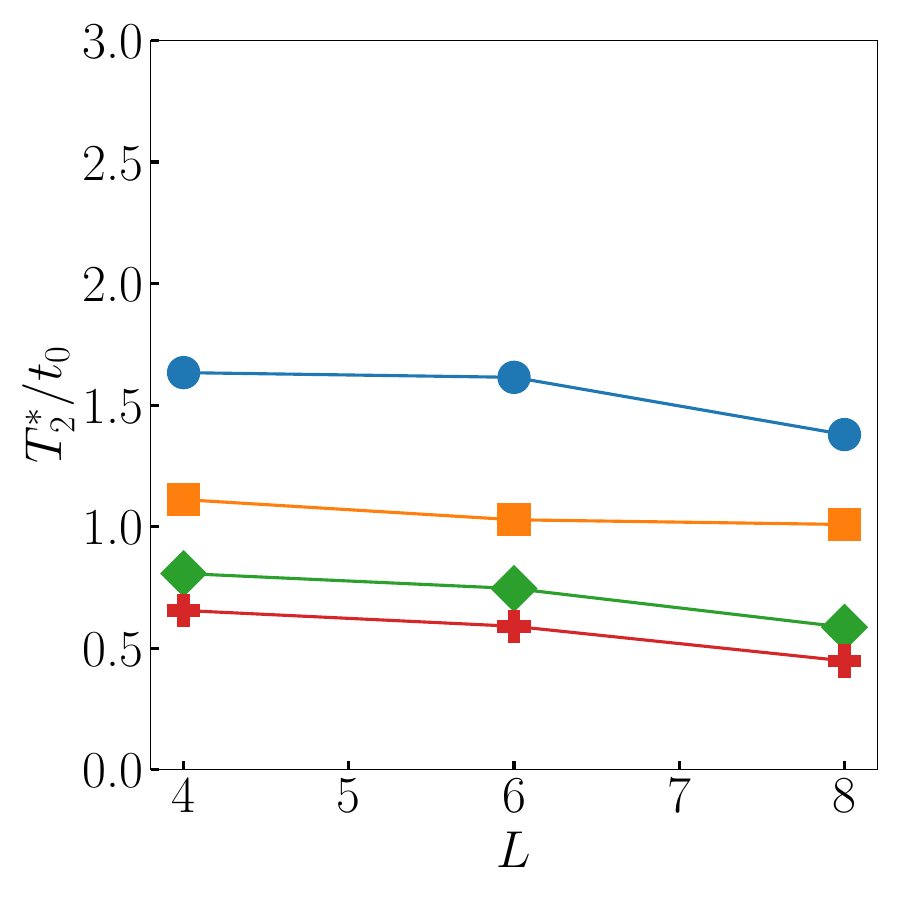}
		\label{fig:subfig_c}}
	\hfill
	\setcounter{subfigure}{3}(d){
		\includegraphics[width=0.42\linewidth]{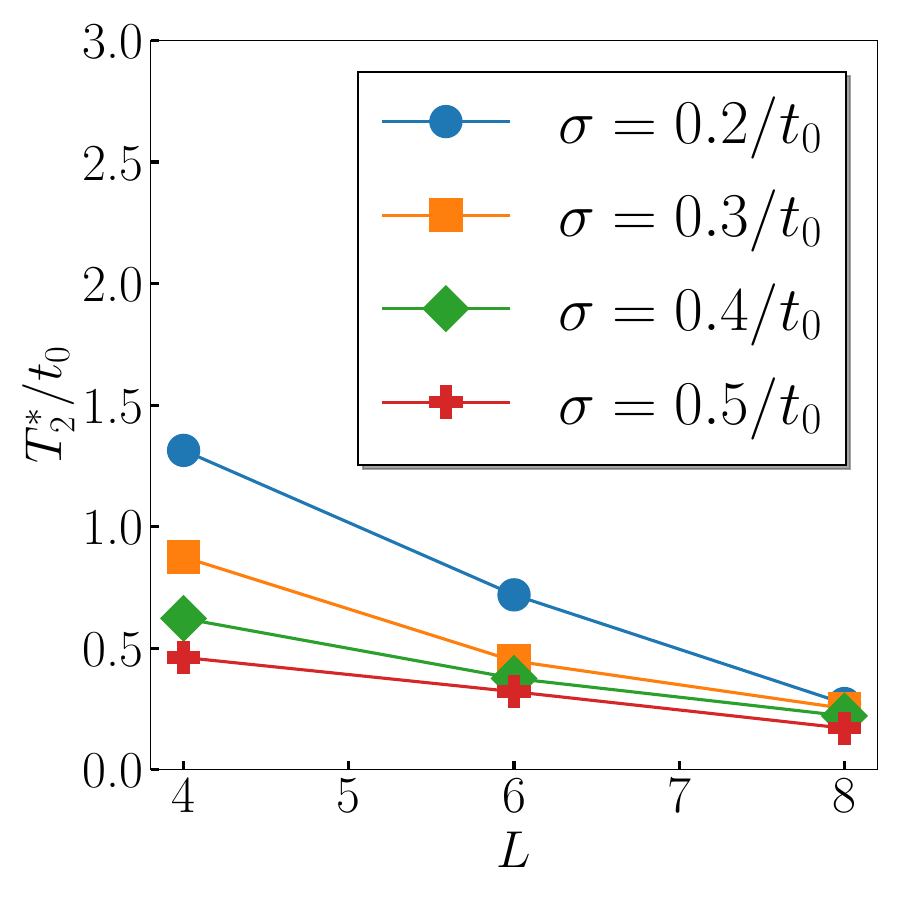}
		\label{fig:subfig_d}}
	\caption{The decoherence time $T_2^{*}$ as a function of system size $L$ for (a) elementary units ``stick'' with ring configuration , for (b) elementary units ``stick'' with chain configuration, for (c) elementary units ``triangle'' with ring configuration, and for (d) elementary units ``triangle'' with chain configuration with different deviation $\sigma$ and interacting strength $ J_0 =100/t_0 $.}
	\label{fig:BCsigma}
\end{figure}

In Fig.~\ref{fig:BCsigma}(a) and Fig.~\ref{fig:BCsigma}(b), we present the decoherence times for the elementary units ``stick'' with ring and chain configuration, considering different deviation $\sigma$ and interacting strength $ J_0 =100/t_0 $. Due to the limitation of the elementary unit ``stick'', the case of $L=4$ is excluded as it consists of only two elementary units, which cannot form a ring. Therefore, we focus on the $L=6$ case for the ring. Notably, in the ring configuration, the overall decoherence times are longer compared to the chain. 
The dependence of the ring and chain to system size varies with different $\sigma$. For relatively large $\sigma$ values, such as $\sigma=0.5/t_0$, the decoherence time for the ring configuration decreases faster as the system grows. However, for relatively small $\sigma$ values, e.g. $\sigma=0.2/t_0$, the decoherence time decreases slower as the system size increases. 
Nevertheless, regardless of the value of $\sigma$, the stability of the ring is consistently better to that of the chain at the same system size.

In Figures~\ref{fig:BCsigma}(c) and~\ref{fig:BCsigma}(d), we investigate the decoherence times of the elementary unit ``triangle'' for both ring and chain configurations, taking into account varying values of deviation $\sigma$ and interaction strength $J_0 = 100/t_0$. Similar to the ``stick'' configuration, the ring configuration consistently shows longer overall decoherence times compared to the chain. Unlike the ``stick'' configuration, however, the ring demonstrates weaker dependence to system size for the elementary unit ``triangle". This implies that, irrespective of the $\sigma$ value, the decoherence time in the ring configuration decreases at a slower rate as the system size increases. Nonetheless, as observed in earlier comparisons, the ring configuration continues to outperform the chain in terms of stability, maintaining longer decoherence times than the chain at the same system size.


Another observation from Fig.~\ref{fig:BCsigma} is that the ``stick'' elementary unit consistently outperforms the ``triangle'' in terms of stability. For example, by comparing Fig.~\ref{fig:BCsigma}(a) and Fig.~\ref{fig:BCsigma}(c), or Fig.~\ref{fig:BCsigma}(b) and Fig.~\ref{fig:BCsigma}(d), we can clearly see that, under the same $\sigma$ and $L$ with the same geometric configuration, the decoherence time of the quantum state with the ``stick'' elementary unit is always longer than that with the ``triangle'' elementary unit. We conclude that the ``stick'' elementary unit displays greater stability compared to the ``triangle'' unit.

The stability of different qubit configurations under noise can be influenced by various factors, but we propose that symmetry plays a crucial role in the superior stability of ring configurations compared to chains. Taking a small system with $L=3$ as an example \cite{Buterakos2021}, the Hamiltonian for the ring configuration commutes with the total spin, meaning that the total spin is conserved. This conservation provides a degree of protection against decoherence, preserving the quantum state. Chain and tree configurations do not exhibit this property. As the size of the system increases, the total spin conservation is no longer maintained in the ring configuration. However, a residual protective effect from the initial symmetry persists. We attribute this lingering protection to the inherent symmetry of the ring configuration, which we believe is the primary reason for its enhanced stability compared to chains.

\subsection{Entanglement entropy for different geometric connectivity}
\label{subsec:result2}

Due to the potential for large system sizes in tree configurations, our current study of the tree configuration specifically focuses on the results of the ``node'' elementary unit for a system size of $L=8$. In addition to extracting the decoherence lifetime, we also calculate the entanglement entropy, a highly relevant physical quantity in theoretical studies. We partition the system into a target system $\alpha$ and an environment $\beta$, and obtain the density matrix of the target system by performing a partial trace operation: $\rho_{\alpha}(t)={\rm Tr}_{\beta}\rho(t)$, where $\rho(t)=|\Phi(t)\rangle\langle\Phi(t)|$ represents the density matrix of the entire system at time $t$. The entanglement entropy is defined as
\begin{equation}
	S(t) = -{\rm Tr}_{\alpha} \left[\rho_{\alpha}(t) \ln \rho_{\alpha}(t)\right],
\end{equation}
where $ \rho(t)=|\Phi(t)\rangle\langle\Phi(t)| $ is the density matrix at time $ t $. 

We investigate the influence of varying the deviation $\sigma$ on the decoherence time for a system with $L=8$ qubits. In the case of the ``node'' elementary unit, as $\sigma$ increases, the decoherence time gradually decreases for all three connectivity types: ring, chain, and tree. However, the ring configuration shows a faster rate of decrease compared to the chain and tree. Despite the variations, the ring consistently outperforms the chain and tree in terms of decoherence time, indicating that the ring configuration is more stable, in line with our previous conclusions.

\begin{figure}[t]
	\includegraphics[width=0.9\linewidth]{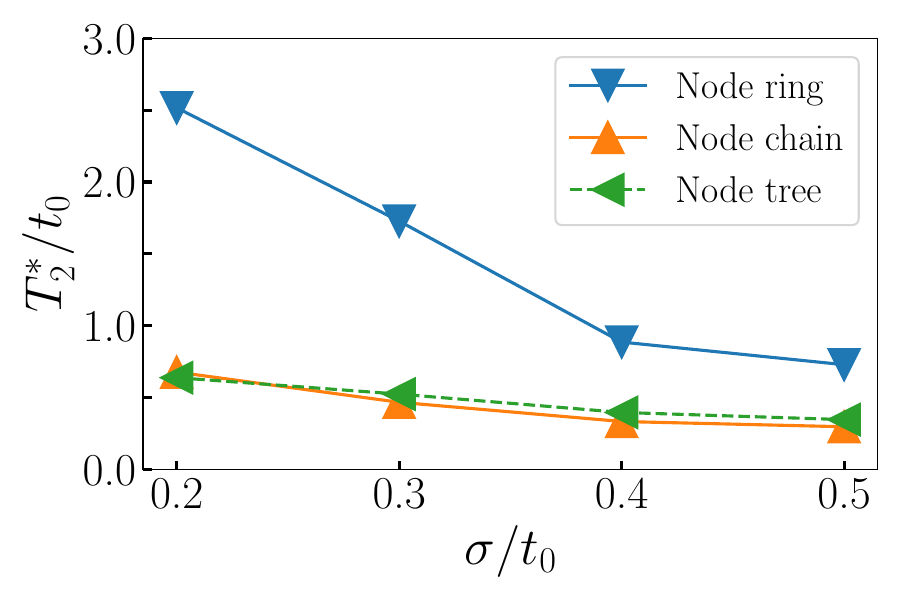}
	\caption{
			The decoherence time $ T_2^{*} $ as a function of $ \sigma $ for elementary units ``node'' for ring, chain and tree configuration at system size $ L=8 $. The interacting strength $ J_0=100/t_0 $. 
			\label{fig:tree}
		}
\end{figure}

Moreover, an important observation from the figures is that the chain and tree configurations exhibit similar decoherence times (Fig.~\ref{fig:tree}). This finding highlights the comparable stability of these two connectivity types. Our results help in strengthening our understanding of how different connectivity patterns impact the overall stability and performance of quantum systems. The geometric configuration plays a crucial role in determining the decoherence behavior, which is valuable for designing robust quantum systems and devising quantum error correction strategies.

\begin{figure}[t]
	\centering
	\setcounter {subfigure} {0} (a){
		\includegraphics[width=0.7\linewidth]{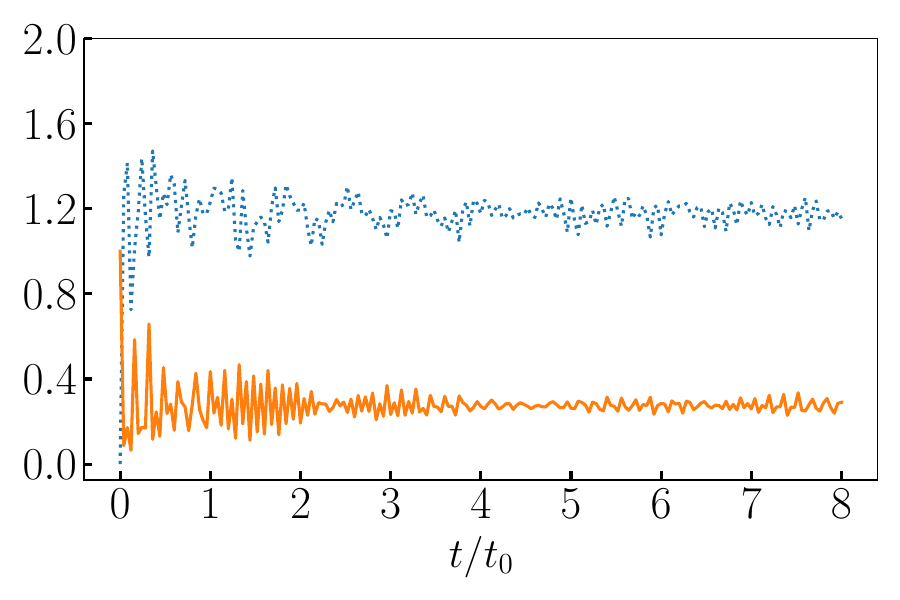}
	}

	\setcounter {subfigure} {0} (b){
		\includegraphics[width=0.7\linewidth]{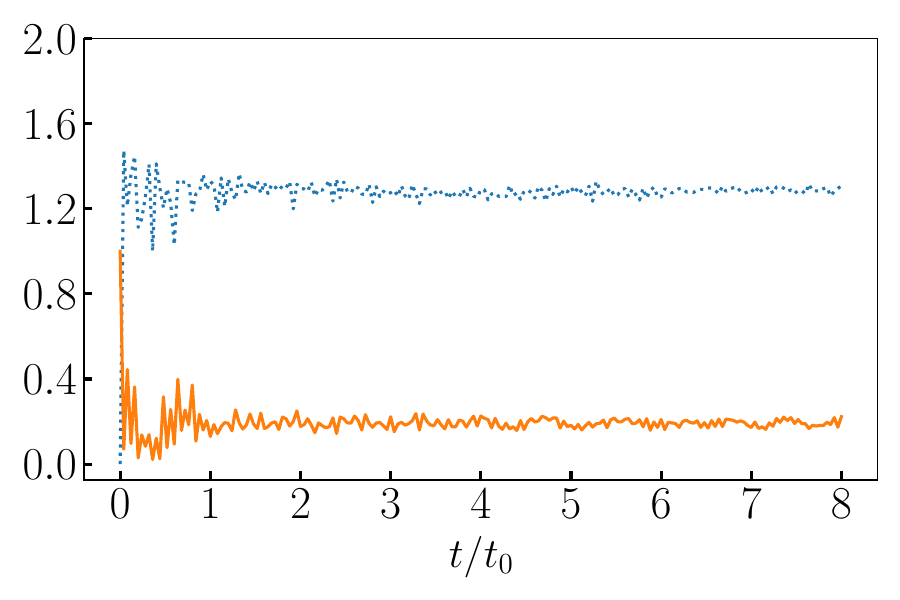}}
	
	\setcounter {subfigure} {0} (c){
		\includegraphics[width=0.7\linewidth]{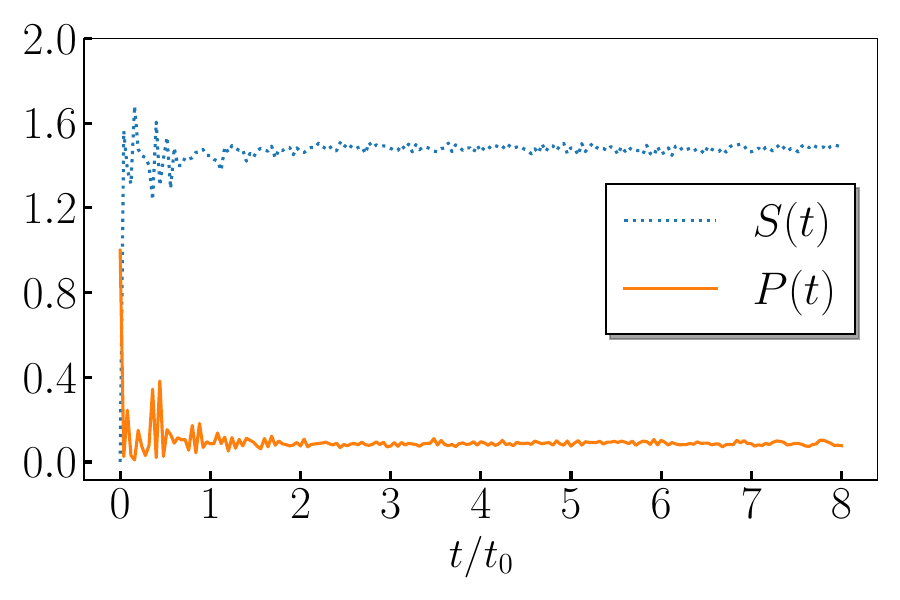}
	}

	\caption{The entanglement entropy $ S(t) $ \blue{(upper curves) and the return probability $P(t)$ (lower curves)}  for elementary unit ``node'' for (a) ring, (b) chain and (c) tree configuration at system size $ L=8 $. The trend of the entanglement entropy $ S(t) $ is very close to that of the return probability $ P(t)$, so both can be used to characterize the decoherence lifetime.}
	\label{fig:entropy}
\end{figure}

In Fig.~\ref{fig:entropy}, we observe a remarkable similarity between the trends of the entanglement entropy and the return probability $P(t)$, indicating that both measures can effectively describe the decoherence time of the system. This finding highlights the potential of utilizing the entanglement entropy as an alternative measure to characterize the degradation of quantum coherence. Additionally, our analysis demonstrates that the stability of the tree configuration configuration is comparable to that of the chain configuration, regardless of whether it is assessed based on the entanglement entropy or the decoherence time. This observation suggests that both chain and tree connectivities exhibit similar levels of vulnerability to decoherence effects, further emphasizing the importance of geometric connectivity in determining the overall stability of quantum systems. These insights contribute to advancing our understanding of the interplay between connectivity patterns and decoherence dynamics, paving the way for improved strategies in quantum information processing and quantum device design.

\section{Conclusions and discussions}
\label{conclusion}
In summary, this paper investigates the influence of connectivity on the decoherence time under quasi-static Heisenberg noise. By examining various configurations of elementary units, including rings, chains, and trees, we uncover interesting and counterintuitive results. Specifically, we find that rings exhibit greater stability compared to chains, contrary to the expectation that more links between qubits leads to decreased stability. We attribute this difference to the role of symmetry, which can protect the quantum state from leakage into other subspaces. 

Additionally, we observe that the ``stick'' configuration is more stable than the ``triangle'' configuration, which is expected due to the disparity in their average connectivity. Notably, the disparity in decoherence time between rings and chains is significant, even though their average connectivity is relatively similar. These findings shed light on the complex interplay between connectivity and stability in quantum systems.

Furthermore, our results highlight the comparable stability of tree and chain configurations, as evident from both the entanglement entropy and decoherence time analyses. This insight underscores the importance of considering connectivity patterns when designing quantum systems and devising quantum error correction strategies. Overall, our findings should help in strengthening our understanding of the impact of connectivity on decoherence dynamics, providing inspirations for the development of robust quantum technologies.
\acknowledgments
This work is supported by the Key-Area Research and Development Program of GuangDong Province  (Grant No. 2018B030326001) and the Research Grants Council of Hong Kong (Grant No. CityU 11304920).
\appendix

\section{The effect of single-qubit term}\label{app:single}

\begin{figure}[t]
	\includegraphics[width=0.9\linewidth]{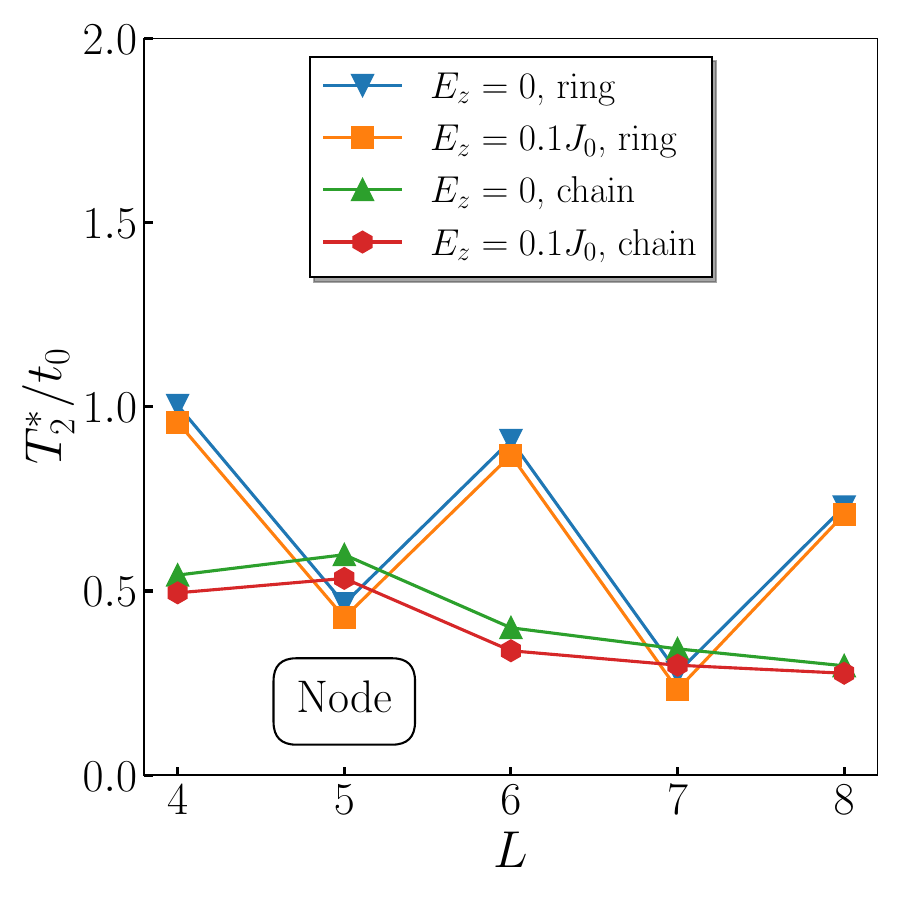}
	\caption{
		\blue{
		The decoherence time $T_2^{*}$ is plotted as a function of $ L $ for systems composed of nodes in rings and chains, both with and without single-qubit terms. We take $\sigma=0.4/t_0$, $J_0=100/t_0$ and $ E_{z}=0.1J_{0} $.}
		\label{fig:single}
	}
\end{figure}

\blue{ In Sec.~\ref{Model and method}, we mentioned that the results shown in the main text are obtained with single-qubit interaction terms (Zeeman terms such as $ E_{z}\sigma_{iz}$) ignored. The main reason is that errors incurred in two-qubit gates are much higher than those for single-qubit gates, as pointed out in \cite{Johnstun2021}, and crosstalk plays a key role in the two-qubit decoherence process \cite{Robert}. In this section, we conduct calculations in presence of single-qubit terms. In a study \cite{Bartolomeo2023} on IBM's Qiskit, the simulated $T_2^{*}$ under single-qubit gate operations is ten times longer than that of two-qubit gate operations. This gives us a rough estimate of $E_{z}\approx0.1J_{0}$. The calculations with $E_{z}=0.1J_{0} $ is shown in Fig.~\ref{fig:single}, which is compared to the results with $E_{z}=0$. The results from   $E_{z}=0.1J_{0} $ and $E_{z}=0$ are rather close, suggesting that our main conclusion remain valid in presence of a resonable low level of single-qubit error terms.}


\section{Dynamic noise}\label{app:3}
For dynamical noise, we consider the $ 1/f^{\alpha} $ noise with $ \alpha\in[1,3] $, and the time series of the coupling strength $ J_{ij}(t) $ is constructed using the fractional Brownian motion (fBm) method \cite{Yang2016}. 
\begin{figure}[t]
	\centering
	\setcounter {subfigure} {0} (a){
				\label{fig:dynamic_ring}
		\includegraphics[width=0.42\linewidth]{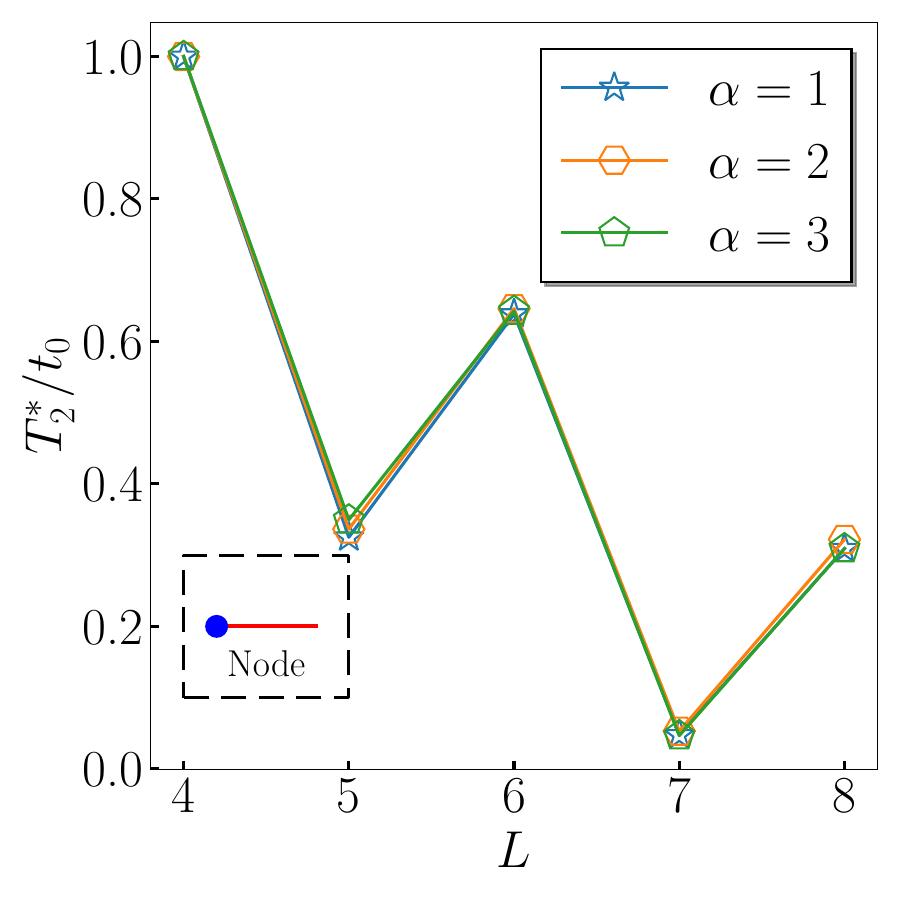}
	}\setcounter {subfigure} {0} (b){
				\label{fig:dynamic_line}
		\includegraphics[width=0.42\linewidth]{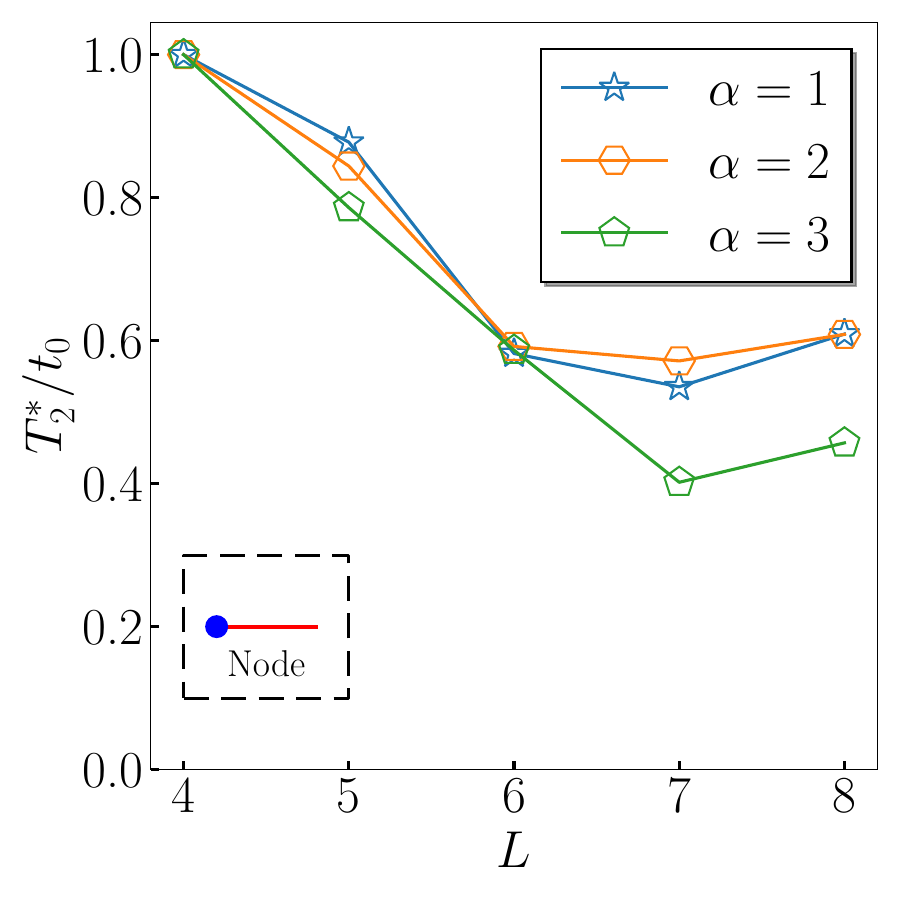}}
	
	\caption{ Variation of the decoherence time with the system size $L$ for (a) ring and (b) chain configuration. Each plot represents different values of $\alpha$ ranging from 1 to 3. The decoherence times are rescaled by dividing each value by the maximum $T_2^{*}$ for comparison. We take $\sigma=0.5/t_0$ and $J_0=100/t_0$. The elementary unit employed in the study is the ``node''. Despite variations in system size and different values of $\alpha$, the decoherence times remain relatively consistent within each geometric connectivity.}
	\label{fig:dynamic}
\end{figure}
\begin{figure}[h]
	\includegraphics[width=0.9\linewidth]{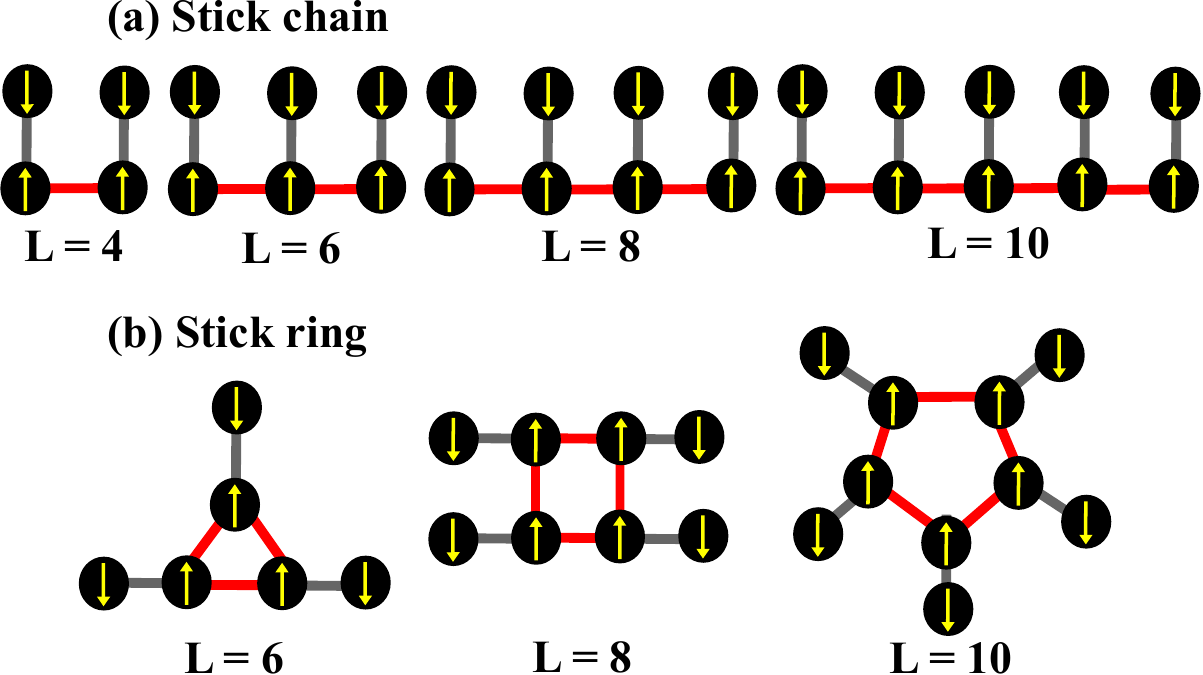}
	\caption{
Visualization of (a) chain and (b) ring configurations of the elementary unit ``stick''. The vertices represent qubits, with gray lines indicating connections within each unit and red lines showing connections between repeated units. The arrows inside the vertices represent the chosen initial states: an upward arrow signifies a spin-up state, while a downward arrow signifies a spin-down state.
		\label{fig:connectivity_stick}
	}
\end{figure}
\begin{figure}[h]
	\includegraphics[width=0.9\linewidth]{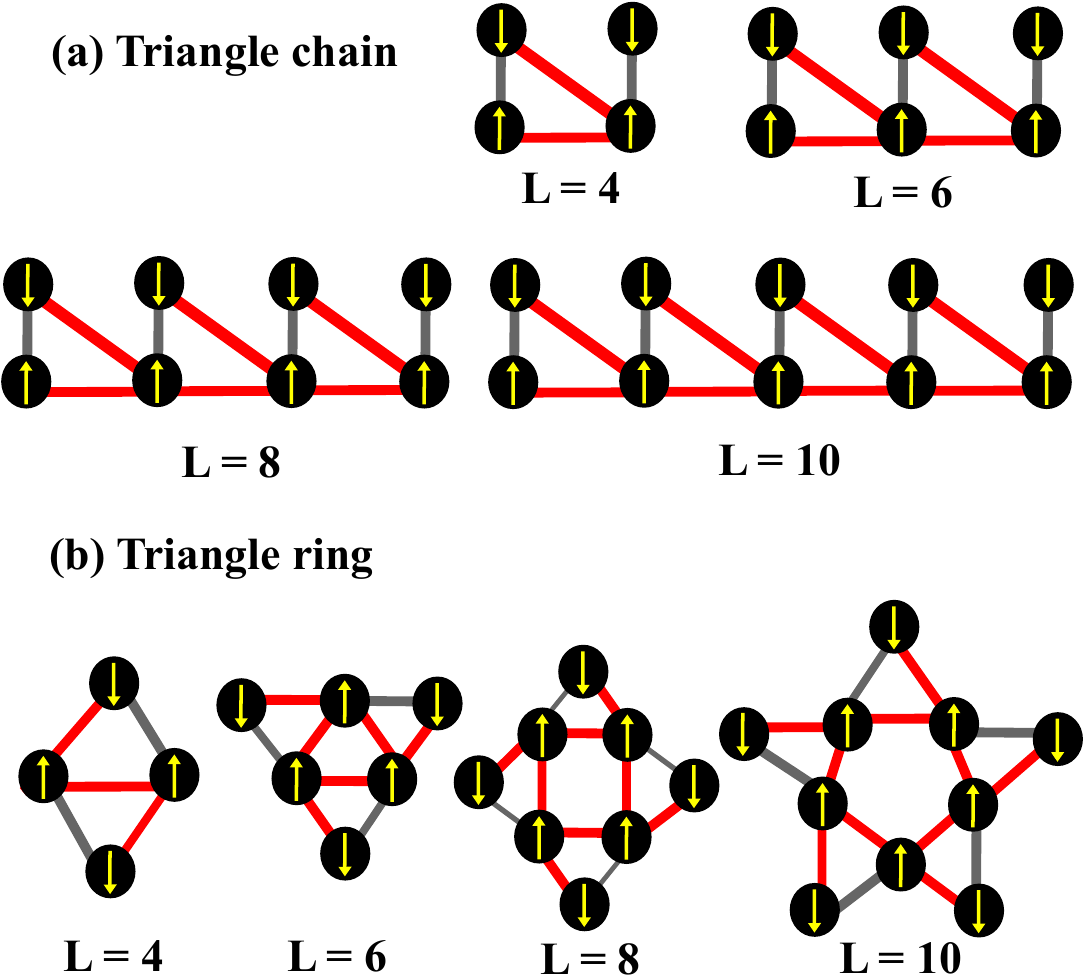}
	\caption{
		Visualization of (a) chain and (b) ring configurations of elementary unit ``triangle''.
		\label{fig:connectivity_tri}
	}
\end{figure}
We examine the variation of the decoherence time with the system size $L$ for two different geometric connectivities: ring and chain. Fig.~\ref{fig:dynamic} shows the results for the ring and chain configurations. Each plot represents different values of $\alpha$, ranging from 1 to 3.

We rescale the decoherence times by dividing each value by the maximum $T_2^{*}$ to facilitate comparison. The chosen value of $\sigma$ is 0.5, and the interacting strength is set as $J_0=100/t_0$. The elementary unit employed in our study is the ``node''. It is noteworthy that, despite variations in the system size and different values of $\alpha$, the decoherence times remain relatively consistent within each geometric connectivity. This observation suggests that the chosen geometric connectivity has a more significant impact on the decoherence time than the specific system size or parameter $\alpha$. 

Remarkably, irrespective of the system size and different values of $\alpha$, the decoherence times exhibit a consistent pattern within each specific geometric connectivity. This finding underscores the significance of the chosen geometric connectivity in determining the decoherence time dynamics.
\section{Visualization of chain and ring configurations studied in this work}\label{app:1}

In this appendix, we present visualizations of chain and ring configurations formed by elementary units ``stick'' and ``triangle,'' with the number of qubits, 
$L$, ranging from 4 to 10.  Fig.~\ref{fig:connectivity_stick} illustrates the chain and ring configurations of ``stick,'' while Fig.~\ref{fig:connectivity_tri} depicts the chain and ring configurations of ``triangle."



\section{Decoherence of GHZ states}\label{app:2}

\begin{figure}[h]
	\includegraphics[width=0.9\linewidth]{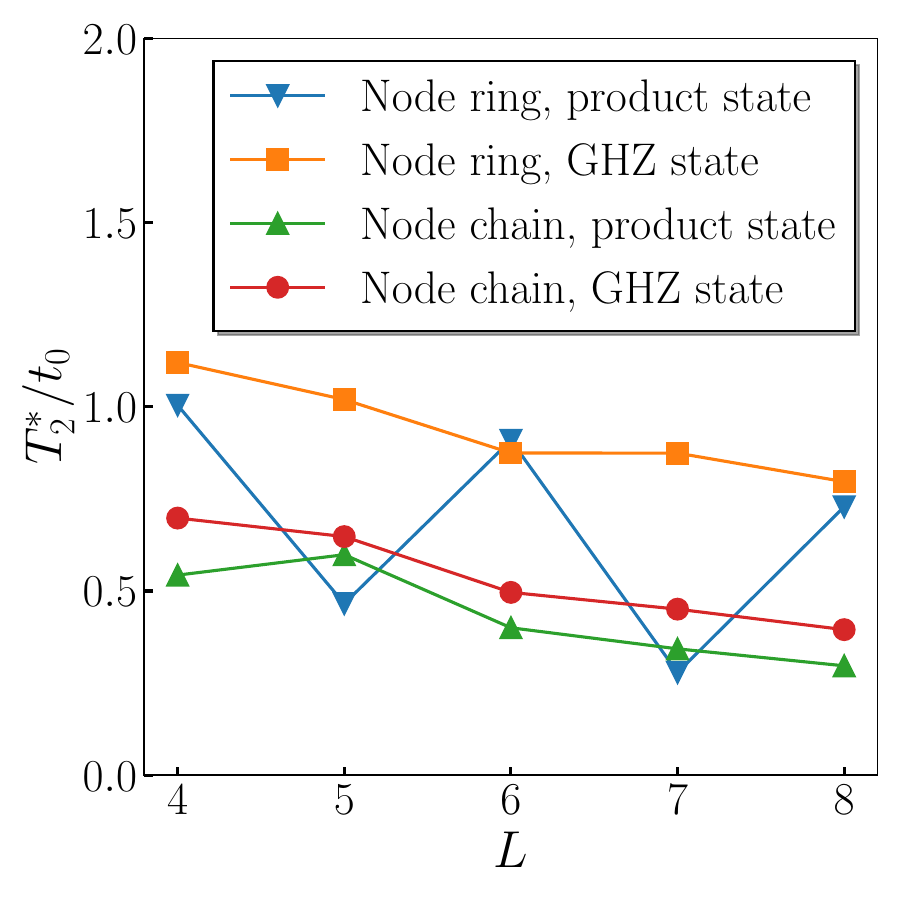}
	\caption{
		The decoherence time $ T_2^{*} $ as a function of $ L $ for the product state and the general GHZ state in different systems.
		\label{fig:GHZ}
	}
\end{figure}

As discussed in the main text, comparing different geometric configurations constrains our choice of initial states. We have chosen the product state 
$ | \Psi(0) \rangle = |\uparrow\downarrow\cdots\rangle $  as one candidate. Another potential choice is the GHZ state, defined as $ | \Psi(0) \rangle_{G} = \frac{1}{\sqrt{2}}(|\uparrow\uparrow\cdots\rangle+|\downarrow\downarrow\cdots\rangle) $. We have conducted calculations for the GHZ state in both ring and chain configurations based on nodes. The results are presented in Fig.~\ref{fig:GHZ}. It is evident that the ring configuration with nodes exhibits a longer decoherence time than the chain configuration, supporting our main conclusion. Additionally, we note the absence of dips for the ring case at odd $L$, indicating that the ``even-odd'' effect does not apply to GHZ states.

Finally, we acknowledge that our choice of initial states is not exhaustive. As near-term quantum devices continue to scale up, the interplay between geometric configuration, initial states, and device properties, including decoherence times, emerges as a significant area of interest that merits further investigation.

\bibliographystyle{apsrev4-1}
%
%

\end{document}